\documentclass[floatfix,showpacs,aps,prb,twocolumn]{revtex4}
\usepackage[utf8]{inputenc}
\usepackage[T1]{fontenc}
\usepackage{bm}
\usepackage{makeidx}
\usepackage{eurosym}
\usepackage{graphicx}
\usepackage{graphics}
\usepackage{amssymb}
\usepackage{amsmath}
\usepackage{color}
\usepackage{float}
\usepackage{hyperref}
\usepackage{fancyhdr}
\usepackage{booktabs}
\usepackage{stackengine}

\DeclareUnicodeCharacter{03B3}{$\gamma$}

\begin{document}

\title{Field-Induced Boson Insulating States in a 2D Superconducting
Electron Gas with Strong Spin-Orbit Scatterings }
\author{Tsofar Maniv}
\author{Vladimir Zhuravlev}
\affiliation{Schulich Faculty of Chemistry, Technion-Israel Institute of Technology,
Haifa 32000, Israel}
\email{maniv@technion.ac.il}
\date{\today }

\begin{abstract}
The phenomenon of field-induced superconductor-to-insulator transitions
observed experimentally in electron-doped SrTiO$_{3}$/LaAlO$_{3}$
interfaces, analyzed recently by menas of 2D superconducting fluctuations
theory (Phys. Rev. B \textbf{104}, 054503 (2021)), is revisited with new
insights associating it with the appearnace at low temperatures of
field-induced boson insulating states. Within the framework of the
time-dependent Ginzburg-Landau approach, we pinpoint the origin of these
states in field-induced extreme softening of fluctuation modes over a large
region in momentum space, upon diminishing temperature, which drives
Cooper-pair fluctuations to condense into mesoscopic puddles in real space.
Dynamical quantum tunneling of Cooper-pair fluctuations out of these
puddles, introduced within a phenomenological approach, which break into
mobile single-electron states, contains the high-field resistance onset
predicted by the exclusive boson theory.
\end{abstract}

\maketitle

\section{\protect\bigskip Introduction}

In a recent paper \cite{MZPRB2021} we have shown that Cooper-pair
fluctuations in a 2D electron gas with strong spin-orbit scatterings can
lead at low temperatures to pronounced magnetoresistance (MR) peaks above a
crossover field to superconductivity. The model was applied to the high
mobility electron systems formed in the electron-doped interfaces between
two insulating perovskite oxides---SrTiO$_{3}$ and LaAlO$_{3}$ \cite%
{Ohtomo04},\cite{Caviglia08}, showing good quantitative agreement with a
large body of experimental sheet-resistance data obtained under varying gate
voltage \cite{Mograbi19}.

The model employed was based on the opposing effects generated by
fluctuations in the superconducting (SC) order parameter: The nearly
singular enhancement of conductivity (paraconductivity) due to fluctuating
Cooper pairs below the nominal (mean-field) critical magnetic field, on one
hand, and the suppression of conductivity, associated with the loss of
unpaired electrons due to Cooper pairs formation, on the other hand. The
self-consistent treatment of the interaction between fluctuations \cite%
{UllDor90},\cite{UllDor91}, employed in these calculations, avoids the
critical divergence of both the Aslamazov-Larkin (AL) paraconductivity \cite%
{AL68} and the DOS conductivity \cite{LV05}, allowing to extend the theory
to regions well below the nominal critical SC transition. The absence of
long range phase coherence implied by this approach is consistent with the
lack of the ultimate zero-resistance state in the entire data analyzed there.

The most intriguing question arising from the Cooper-pair fluctuations
scenario of the superconductor--insulator transition (SIT) presented in Ref.%
\cite{MZPRB2021}, is how Cooper-pairs liquid, whose condensation (in
momentum space) is customarily associated with superconductivity, could
metamorphose into an insulator just by lowering its temperature under
sufficiently high magnetic field ?

For answering this intriguing question we note our use of the time-dependent
Ginzburg-Landau (TDGL) functional approach in consistently evaluating the AL
and the DOS conductivities. Within this exclusive boson approach we have
found in Ref.\cite{MZPRB2021} that at low temperatures the (negative) DOS
conductivity prevails over the AL paraconductivity at fields that roughly
indicate the presence of the observed enhanced MR. Dynamical quantum
tunneling of Cooper-pair fluctuations out of mesoscopic puddles has been
introduced into the theory within a complementary phenomenological approach,
including the contribution of unpaired normal electron states, to account
for the observed experimental data.

In the present paper we reveal the underlying origin of these
low-temperature field-induced boson insulating states by exploiting a
detailed analytical scheme within the framework of the TDGL functional
approach. It is found that strong field-induced suppression of the
fluctuation stiffness parameter at low temperatures resulting in extreme
softening of fluctuating modes over a large region in momentum space,
dramatically enhances the Cooper-pair fluctuations density in mesoscopic
puddles of real space. The resulting large enhancement of the (negative) DOS
conductivity versus the diminishing AL paraconductivity, associated with the
fluctuation mass enhancement, trigger the appearance of insulating states at
high field. Our detailed analysis has also illuminated the mechanism in
which the exclusive fluctuation boson picture is modified within a unified
phenomenological approach. It allows field-induced pair-breaking processes
to develop during dynamical quantum tunneling of Cooper-pair fluctuations
out of mesoscopic puddles, which result in free exchange between the systems
of charge-bosons and unpaired free electrons.

\section{The TDGL functional approach}

The TDGL functional $\mathfrak{L}\left( \Delta ,\mathbf{A}\right) $ of the
order parameter $\Delta \left( \boldsymbol{r},t\right) $ and vector
potential $\mathbf{A}\left( \boldsymbol{r},t\right) $ determines the
Cooper-pairs current density \cite{FuldeMaki70}: 
\begin{equation}
\mathbf{j}\left( \mathbf{r},t\right) =\frac{\partial \mathfrak{L}\left(
\Delta \left( \mathbf{r},t\right) ,\mathbf{A}\left( \mathbf{r},t\right)
\right) }{\partial \mathbf{A}\left( \mathbf{r},t\right) }  \label{j(r,t)}
\end{equation}%
responsible for the AL paraconductivity. In this approach the entire
underlying information about the thin film of pairing electrons system
(which includes in-plane spin-orbit scatterings, Zeeman spin splitting as
well as out-of-plane diamagnetic energy \cite{MZPRB2021}) is incorporated in
the inverse fluctuation propagator (in wavevector-frequency representation) $%
D^{-1}\left( \mathbf{q+}2e\mathbf{A/\hbar },\omega \right) $, mediating
between the order parameter and the GL functional. In the Gaussian
approximation the relation is quadratic, i.e.: 
\begin{eqnarray}
\mathfrak{L}\left( \Delta ,\mathbf{A}\right) &=&\left( \frac{1}{2\pi }%
\right) ^{2}d^{-1}\int d^{2}q\left( \frac{1}{2\pi }\right)  \label{GLLag} \\
&&\times \int d\Omega \left\vert \Delta \left( \mathbf{q},\Omega \right)
\right\vert ^{2}D^{-1}\left( \mathbf{q+}2e\mathbf{A/\hbar },\Omega \right) 
\notag
\end{eqnarray}%
so that the coupling to the external electromagnetic field takes place
directly through the vertex of the Cooper-pair current, defined in Eq.(\ref%
{j(r,t)}).

The corresponding AL time-ordered current-current correlator is given by: 
\begin{eqnarray}
&&Q_{AL}\left( i\Omega _{\nu }\right)=\left( 4eN_{2D}D\right)
^{2}d^{-1}\left( \frac{1}{2\pi }\right) ^{2}\int d^{2}qq_{x}^{2}k_{B}T
\label{Q_ALgen} \\
&&\sum\limits_{\mu =-\infty }^{\infty }C\left( q,\Omega _{\mu }+\Omega _{\nu
}\right) D\left( q,\Omega _{\mu }+\Omega _{\nu }\right) C\left( q,\Omega
_{\mu }\right) D\left( q,\Omega _{\mu }\right)  \notag
\end{eqnarray}%
where $\Omega _{\mu }=2\mu k_{B}T/\hbar ,\Omega _{\nu }=2\nu k_{B}T/\hbar ,$ 
$\mu =0,\pm 1,\pm 2,...,$\ \ \ \ $\nu =0,1,2,....$ are bosonic Matsubara
frequencies, $d$ is the thickness of the detected film, and $N_{2D}=m^{\ast
}/2\pi \hbar ^{2}$ is the single-electron DOS, with an effective mass $%
m^{\ast }$. Here the electrical current is generated along the $x$ axis, $%
q_{z},q_{x}$ are the fluctuation (in-plane) wave-vector components along the
magnetic and electric field directions, respectively, and $q^{2}\equiv
q_{z}^{2}+q_{x}^{2}$.

Explicitly for the model of spin-orbit scatterings employed, the fluctuation
propagator $D\left( q,\Omega _{\mu }\right) $ and its corresponding
effective current vertex $C\left( q,\Omega _{\mu }\right) $ are given by 
\cite{MZPRB2021}:\ 

\begin{eqnarray}
D\left( q,\Omega _{\mu }\right) &=&\frac{1}{N_{2D}\Phi \left( x+\left\vert
\mu \right\vert ;\varepsilon _{H}\right) },  \notag \\
C\left( q,\Omega _{\mu }\right) &=& \frac{1}{4\pi k_{B}T}\Phi ^{\prime
}\left( x+\left\vert \mu \right\vert ;\varepsilon _{H}\right)  \label{D&C}
\end{eqnarray}%
where 
\begin{eqnarray}
&&\Phi \left( x+\left\vert \mu \right\vert ;\varepsilon _{H}\right) =
\varepsilon _{H}+  \notag \\
&& 
\begin{array}{c}
a_{+}\left[ \psi \left( 1/2+f_{-}+x+\left\vert \mu \right\vert \right) -\psi
\left( 1/2+f_{-}\right) \right] \\ 
+a_{-}\psi \left[ \left( 1/2+f_{+}+x+\left\vert \mu \right\vert \right)
-\psi \left( 1/2+f_{+}\right) \right]%
\end{array}
\label{Phi(x,mu)}
\end{eqnarray}%
and: 
\begin{equation}
\varepsilon _{H}\equiv \ln \left( \frac{T}{T_{c0}}\right) +a_{+}\psi \left( 
\frac{1}{2}+f_{-}\right) +a_{-}\psi \left( \frac{1}{2}+f_{+}\right) -\psi
\left( 1/2\right)  \label{eps_H}
\end{equation}

Here $T_{c0}$ is the mean-field SC transition temperature at zero magnetic
field, $\psi $ is the digamma function, $x=\hbar Dq^{2}/4\pi k_{B}T$, where $%
D\equiv \tau _{SO}E_{F}/m^{\ast }$ is the electron diffusion coefficient, $%
E_{F}$- the Fermi energy, and $\varepsilon _{SO}=\hbar /\tau _{SO}$ is the
spin-orbit energy. The system parameters: $\ f_{\pm }=\delta H^{2}+\beta \pm 
\sqrt{\beta ^{2}-\mu ^{2}H^{2}}$, $a_{\pm }=\left( 1\pm \beta /\sqrt{\beta
^{2}-\mu ^{2}H^{2}}\right) /2$ are dimensionless functions of the magnetic
field $H$, with the basic parameters: $\ \beta \equiv \varepsilon _{SO}/4\pi
k_{B}T,$ $\delta \equiv D\left( de\right) ^{2}/2\pi k_{B}T\hslash $, $\mu
\equiv \mu _{B}/2\pi k_{B}T\ $and $\mu _{B}$ the Bohr magneton.

The DOS conductivity is obtained within this TDGL functional approach by
exploiting the Drude formula $\sigma _{DOS}=-2n_{s}e^{2}\tau _{SO}/m^{\ast }$%
, through the Cooper-pair fluctuations density $n_{s}$ \cite{LV05}:%
\begin{equation}
n_{s}=\frac{1}{d}\frac{1}{\left( 2\pi \right) ^{2}}\int \left\langle
\left\vert \phi \left( q\right) \right\vert ^{2}\right\rangle d^{2}q
\label{n_s}
\end{equation}%
with the Cooper-pair momentum distribution function $\left\langle \left\vert
\phi \left( q\right) \right\vert ^{2}\right\rangle $ derived by exploiting
the frequency-dependent GL functional, Eq.(\ref{GLLag}). This is done by
rewriting Eq.(\ref{GLLag}) in terms of the frequency and wavenumber
representations GL wave functions $\phi \left( q,\Omega \right) $, after
analytic continuation to real frequencies $i\Omega _{\mu }\rightarrow \Omega 
$, i.e.:

\begin{eqnarray}
\mathfrak{L}\left( \Delta \right) &=&\int \frac{d^{2}q}{\left( 2\pi \right)
^{2}}\int \frac{d\Omega }{2\pi }\left\vert \Delta \left( q,\Omega \right)
\right\vert ^{2}D\left( q,\Omega \right) ^{-1}  \label{Lag(phi)} \\
&=&\int \frac{d^{2}q}{\left( 2\pi \right) ^{2}}\int \frac{d\Omega }{2\pi }%
\left\vert \phi \left( q,\Omega \right) \right\vert ^{2}L\left( q,\Omega
\right) ^{-1}=\mathfrak{L}\left( \phi \right)  \notag
\end{eqnarray}%
where the transformed inverse propagator $L\left( q,\Omega \right) ^{-1}$
given by:

\begin{equation}
N_{2D}D\left( q,\Omega \right) =\mathcal{A}k_{B}TL\left( q,\Omega \right) 
\notag
\end{equation}%
and: $\mathcal{A}\equiv 4\pi ^{2}k_{B}T/7\zeta \left( 3\right) E_{F}$, with: 
$\zeta \left( 3\right) \simeq 1.202$.

For the sake of clarity of the analysis that follows we expand $L^{-1}\left(
q,\Omega \right) $ to leading orders in small $q$ and $\Omega $, i.e.:

\begin{equation}
L^{-1}\left( q,\Omega \right) \simeq \mathcal{A}k_{B}T\left[ \varepsilon
_{H}+\widetilde{\eta }\left( H\right) \xi ^{2}q^{2}\right] -i\left( \hbar
\Omega \right) \gamma _{GL}  \label{L^-1}
\end{equation}%
where: 
\begin{eqnarray}
\eta \left( H\right) &=&a_{+}\psi ^{\prime }\left( \frac{1}{2}+f_{-}\right)
+a_{-}\psi ^{\prime }\left( \frac{1}{2}+f_{+}\right) ,  \label{eta} \\
\widetilde{\eta }\left( H\right) &\equiv &\frac{\eta \left( H\right) }{\eta
\left( 0\right) }=\frac{2\eta \left( H\right) }{\pi ^{2}}  \notag
\end{eqnarray}%
$\xi =\sqrt{\pi \hbar D/8k_{B}T}$ is the dirty-limit coherence length and $%
\gamma _{GL}=\widetilde{\eta }\left( H\right) \pi \mathcal{A}/8$ is the
dimensionless GL Cooper-pair life time. \ Thus, using Eq.(\ref{L^-1}) the
momentum distribution function is related to the fluctuation propagator
through \cite{LV05}:

\begin{equation*}
\left\langle \left\vert \phi _{q}\right\vert ^{2}\right\rangle
=2k_{B}T\gamma _{GL}\int \frac{d\left( \hbar \Omega \right) }{2\pi }%
\left\vert L\left( q,\Omega \right) \right\vert ^{2}
\end{equation*}%
which readily yields:

\begin{equation}
\left\langle \left\vert \phi _{q}\right\vert ^{2}\right\rangle \simeq \left( 
\frac{7\zeta \left( 3\right) E_{F}}{4\pi ^{2}k_{B}T}\right) \frac{1}{%
\varepsilon _{H}+\eta \left( H\right) \left( \frac{\hbar D}{4\pi k_{B}T}%
\right) q^{2}}  \label{momdistr}
\end{equation}

\bigskip

\section{Conductance fluctuations at very low temperatures}

In order to reveal the origin of the puzzling insulating state that emerges
in our approach from SC fluctuations we will consider in this section the
fluctuations contributions to the sheet conductivity in the magnetic fields
region where they are rigorously derivable from the microscopic Gor'kov's
Ginzburg-Landau theory, i.e. above the nominal (mean-field) critical field,
determined from the vanishing of the Gaussian critical shift-parameter $%
\varepsilon _{H}$ (Eq.(\ref{eps_H})). There are no restrictions on the
temperature $T$ as we are mainly interested in the low temperatures region
well below $T_{c0}$ down to the limit of $T\rightarrow 0$.

\subsection{DOS conductivity}

Using Eq.(\ref{momdistr}) in Eq.(\ref{n_s}) and the Drude formula, the DOS\
conductivity is written in the form: 
\begin{eqnarray}
\sigma _{DOS}d &\simeq &-3.5\zeta \left( 3\right) \left( \frac{G_{0}}{\pi }%
\right) \int_{0}^{x_{c}}\frac{dx}{\varepsilon _{H}+\eta \left( H\right) x}
\label{sig_DOS} \\
&\simeq &-4.2\left( \frac{e^{2}}{\pi ^{2}\hbar }\right) \frac{1}{\eta \left(
H\right) }\ln \left( 1+\frac{\eta \left( H\right) x_{c}}{\varepsilon _{H}}%
\right)  \notag
\end{eqnarray}%
where $G_{0}=e^{2}/\pi \hbar $ is the conductance quantum, $x_{c}=\hbar
Dq_{c}^{2}/4\pi k_{B}T$, with $q_{c}$ the cutoff wave number, and $3.5\zeta
\left( 3\right) \simeq 4.207$. \ It is interesting to compare this result
with the result of the fully microscopic (diagrammatic) approach presented
in Ref.\cite{LV05} for a multilayer of 2D electron systems in the zero field
limit. Using the notation employed in Ref.\cite{LV05} (according to which $%
\hbar =k_{B}=1$ and the distance between layers is $s$) the corresponding
DOS conductivity is given by:

\begin{equation}
\sigma _{DOS}^{LV}=-\kappa \left( T\tau \right) \frac{\pi e^{2}}{2s}\frac{1}{%
\left( 2\pi \right) ^{2}}\int \frac{\eta _{\left( 2\right) }d^{2}q}{%
\varepsilon +\eta _{\left( 2\right) }q^{2}}  \label{sig_DOS^LV}
\end{equation}%
where $\varepsilon \equiv \ln \left( T/T_{c0}\right) $, $\eta _{\left(
2\right) }=\pi D/8T$, and the dirty limit: $\kappa \left( T\tau \right)
_{T\tau \ll 1}\rightarrow 8\times 7\zeta \left( 3\right) /\pi ^{4}$. \ 

Eq.(\ref{sig_DOS^LV}) is in full agreement with the zero-field limit of Eq.(%
\ref{sig_DOS}) derived within our TDGL functional approach. This agreement
is quite remarkable since the coefficient $\kappa \left( T\tau \right)
=8\times 7\zeta \left( 3\right) /\pi ^{4}$ was obtained by summing the
contributions of four diagrams following a lengthy calculation involving
external impurity-scattering renormalization of pair vertices (i.e.
connected to the current vertices by electron lines).

\ \ 

\subsection{Paraconductivity}

The AL contribution to the sheet conductance is calculated by analytically
continuing the time-ordered current-current correlator derived by using Eq.(%
\ref{Q_ALgen}), i.e.: 
\begin{eqnarray}
&&Q_{AL}\left( i\Omega _{\nu }\right) =k_{B}T\left( \frac{2e}{\hbar }\right)
^{2}\left( \frac{1}{2\pi d}\right) \int\limits_{0}^{x_{c}}xdx  \label{Q_AL}
\\
&&\sum\limits_{\mu =0,\pm 1,\pm 2,....}\frac{\Phi ^{\prime }\left(
x+\left\vert \mu +\nu \right\vert ;\varepsilon _{H}\right) }{\Phi \left(
x+\left\vert \mu +\nu \right\vert ;\varepsilon _{H}\right) }\frac{\Phi
^{\prime }\left( x+\left\vert \mu \right\vert ;\varepsilon _{H}\right) }{%
\Phi \left( x+\left\vert \mu \right\vert ;\varepsilon _{H}\right) }  \notag
\end{eqnarray}%
from the imaginary Matsubara frequency $i\Omega _{\nu }$ to the real
frequency $\Omega $ in the static limit, i.e.: $Q_{AL}\left( i\Omega _{\nu
}\right) \rightarrow Q_{AL}^{R}\left( \Omega \right) $; $\sigma
_{AL}=\lim_{\Omega \rightarrow 0}\left( i/\Omega \right) \left[
Q_{AL}^{R}\left( \Omega \right) -Q_{AL}^{R}\left( 0\right) \right] $. \ It
is interesting to note that under direct analytic continuation of the
discrete summation in Eq.(\ref{Q_AL}) about zero frequency, i.e. $\nu
\rightarrow \hbar \Omega /2\pi ik_{B}T\rightarrow 0$, all nonzero
Matsubara-frequency terms are cancelled out and the remaining $\mu =0$ term
can be written in the form: 
\begin{equation}
\sigma _{AL}d=\frac{1}{4}\left( \frac{G_{0}}{\pi }\right)
\int\limits_{0}^{x_{c}}\left( \frac{\Phi ^{\prime }\left( x;\varepsilon
_{H}\right) }{\Phi \left( x;\varepsilon _{H}\right) }\right) ^{2}dx
\label{sig_ALd}
\end{equation}

Exploiting the linear approximation of Eq.(\ref{Phi(x,mu)}), i.e.: \ $\Phi
\left( x;\varepsilon _{H}\right) \simeq \varepsilon _{H}+\eta \left(
H\right) x$, and performing the integration over $x$ analytically we find:

\begin{equation}
\sigma _{AL}d\simeq \left( \frac{e^{2}}{4\pi ^{2}\hbar }\right) \frac{\eta
\left( H\right) }{\varepsilon _{H}\left( 1+\frac{\varepsilon _{H}}{\eta
\left( H\right) x_{c}}\right) }  \label{sig_ALlin}
\end{equation}

Note, that in the zero field limit, where $\eta \left( H\rightarrow 0\right)
=\pi ^{2}/2$ , this result is by a factor of 2 larger than the well-known
result obtained, e.g. in \cite{LV05} by using a fully microscopic
(diagrammatic) approach. The discrepancy is not related to the different
calculational approaches employed but is due to the different schemes of
analytic continuation, used in both approaches, in evaluating the retarded
response function from the time-ordered correlator. The smaller prefactor is
obtained by using the common contour-integration scheme consisting of three
sub-contours (see Ref.\cite{MZAnalCont}). This ambiguity seems to indictate
that the electrical response in the low frequency range is more intricate
than commonly thought and well established in the classic literature. One
may interpret it as, e.g. bistable situation, however the whole issue calls
for further investigation. In any numerical computation performed in this
paper we will adopt the smaller prefactor consistently with the common
analytic continuation scheme. Again, as for the DOS conductivity, the
agreement with the result of the microscopic approach at zero field is quite
remarkable given the fact that in the fully microscopic theory pair vertices
in the AL diagram are renormalized from outside by impurity scattering
ladders between single electron lines.

The exclusive boson TDGL functional approach employed here treats
consistently the DOS and the AL terms as functionals of the fluctuations
propagator whose field dependence exclusively determines the field
dependence of the conductivity.

\subsection{Divergent boson mass at low temperatures}

Combining Eq.(\ref{sig_DOS}) with Eq.(\ref{sig_ALlin}), the resulting
expression for the total fluctuations contributions to the sheet
conductance, $\sigma ^{fluct}d=\sigma _{AL}d+\sigma _{DOS}d$, highlights the
complementary roles played by the stiffness parameter $\eta \left( H\right) $
in the AL and DOS conductivities. The importance of $\eta \left( H\right) $
in controlling the development of an insulating bosonic state at low
temperatures and high magnetic field is clearly revealed by considering the
extreme situation of its zero temperature limit.

To effectively investigate this limiting situation it will be helpful to
rewrite $\eta \left( H\right) $ (see Eq.(\ref{eta})) as a sum over fermionic
Matsubara frequency, that is: 
\begin{equation}
\eta \left( h\right) =\sum\limits_{n=0}^{\infty }\frac{\chi _{n}^{2}-%
\overline{\mu }^{2}h^{2}}{\left[ \chi _{n}\left( \chi _{n}-2\beta \right) +%
\overline{\mu }^{2}h^{2}\right] ^{2}}  \label{eta(h)}
\end{equation}%
where: 
\begin{equation}
\chi _{n}=n+1/2+2\beta +\overline{\delta }h^{2}  \label{chi_n}
\end{equation}%
and: $h\equiv H/H_{c\parallel 0}^{\ast },t\equiv T/T_{c}^{\ast },\overline{%
\mu }=\mu _{0}/t,\mu _{0}\equiv \mu _{B}H_{c\parallel 0}^{\ast }/2\pi
k_{B}T_{c}^{\ast }$, $\ \overline{\delta }=\delta _{0}/t,\delta _{0}\equiv
D\left( deH_{c\parallel 0}^{\ast }\right) ^{2}/2\pi k_{B}T_{c}^{\ast
}\hslash $, with $H_{c\parallel 0}^{\ast }$ and $T_{c}^{\ast }$ being
characteristic scales of the critical parallel magnetic field and critical
temperature, respectively.

At very low temperatures, $t\ll 1$, and finite magnetic field, $h>0$, the
discrete summation in Eq.(\ref{eta(h)}) transforms into integration and:

\begin{equation}
\eta \left( h\right) \rightarrow t\int\limits_{0}^{\infty }d\nu \frac{%
\varkappa _{\nu }^{2}-\mu _{0}^{2}h^{2}}{\left[ \varkappa _{\nu }\left(
\varkappa _{\nu }-2\beta _{0}\right) +\mu _{0}^{2}h^{2}\right] ^{2}}=t\left( 
\frac{\eta _{0}\left( h\right) }{h^{2}}\right)  \label{eta(h)t0}
\end{equation}%
where $\varkappa _{\nu }=\nu +2\beta _{0}+\delta _{0}h^{2}$, 
\begin{equation}
\eta _{0}\left( h\right) \equiv \frac{\delta _{0}h^{2}+2\beta _{0}}{\left(
\delta _{0}h^{2}+2\beta _{0}\right) \delta _{0}+\mu _{0}^{2}}
\label{eta_0(h)}
\end{equation}%
and $\beta _{0}\equiv \varepsilon _{SO}/4\pi k_{B}T_{c}^{\ast }$. \ Note
that at zero magnetic field: $\eta \left( h=0\right) =\sum_{n=0}^{\infty
}\left( n+1/2\right) ^{-2}=\psi ^{\prime }\left( 1/2\right) =\pi ^{2}/2$,
independent of temperature. Thus, the low temperature limit of the sheet
conductance at fields above the nominal critical field $H_{c\parallel
0}^{\ast }$ can be written in the form: 
\begin{align}
& \left( \sigma ^{fluct}\right) _{h>1,t\ll 1}d\rightarrow \left( \frac{G_{0}%
}{\pi }\right) \left[ t\left( \frac{\eta _{0}\left( h\right) }{8h^{2}}%
\right) \frac{1}{\varepsilon _{h}\left( 1+\frac{h^{2}\varepsilon _{h}}{\eta
_{0}\left( h\right) x_{0}}\right) }\right.  \notag \\
& \left. -\frac{1}{t}\left( \frac{3.5\zeta \left( 3\right) h^{2}}{\eta
_{0}\left( h\right) }\right) \ln \left( 1+\frac{\eta _{0}\left( h\right)
x_{0}}{h^{2}\varepsilon _{h}}\right) \right]  \label{sig^fluctdt0}
\end{align}%
where $x_{0}\equiv \hbar Dq_{c}^{2}/4\pi k_{B}T_{c}^{\ast }$ is the
temperature-independent dimensionless cutoff parameter{\normalsize . }Note
the factor of 8 in the denominator of the AL term which follows the common
scheme of analytic continuation, as discussed below Eq.(\ref{sig_ALlin}). It
should be stressed at this point that the temperature-independent argument
of the logarithmic factor in Eq.(\ref{sig^fluctdt0}) (see Ref.\cite%
{footnote2}) is consistent with the temperature-dependent cutoff parameter 
{\normalsize $x_{c}=x_{0}/t$. }It should be also noted here that, despite
the divergence of {\normalsize $x_{c}$ }in the $t\rightarrow 0$\ limit, the
linear approximation $\Phi \left( x;\varepsilon _{H}\right) \simeq
\varepsilon _{H}+\eta \left( H\right) x$ used in deriving Eq.(\ref%
{sig^fluctdt0}) is valid in the entire range of integration below the cutoff 
{\normalsize $x_{c}$ }(see Appendix A).

Thus, we conclude that in the $t\rightarrow 0$\ limit the AL
paraconductivity follows the vanishing stiffness parameter $\eta \left(
h\right) \propto t$, Eq.(\ref{eta(h)t0}), whereas the DOS conductivity
diverges with $1/\eta \left( h\right) \propto 1/t$. Both effects have the
same origin: The divergent effective mass of the fluctuations, which leads
directly to the former effect and indirectly to the latter effect through
extreme softening of the fluctuation modes over a large region in momentum
space, which results in large accumulation of Cooper-pairs within
fluctuation puddles, whose characteristic spatial size (localization length):

\begin{equation}
\widetilde{\xi }\left( t\rightarrow 0\right) =\frac{1}{h}\left( \frac{\eta
_{0}\left( h\right) }{\varepsilon _{h}}\frac{\hbar D}{4\pi k_{B}T_{c}^{\ast }%
}\right) ^{1/2}  \label{xit0}
\end{equation}%
remains finite in this extreme limiting situation. The decreasing asymptotic
field dependence ($\eta \left( h\right) \propto 1/h^{2}$) of the stiffness
parameter (see Eq.(\ref{eta(h)t0})) further enhances the sheet resistance at
high fields by diminishing the localization length ($\widetilde{\xi }\left(
t\rightarrow 0\right) \propto 1/h\sqrt{\varepsilon _{h}}$). \ 

Finally, based on typical values of our fitting parameters (including $%
x_{0}\equiv \hbar Dq_{c}^{2}/4\pi k_{B}T_{c}^{\ast }=0.015$), we use Eq.(\ref%
{xit0}) for determining the value of the cutoff wavenumber $q_{c}$ on the
scale of the inverse temperature independent coherence length $\widetilde{%
\xi }^{-1}\left( t\rightarrow 0\right) $. Thus, at field just above the
"nominal" critical field $H_{c\parallel 0}^{\ast }=4.5$T ($\varepsilon
_{h\gtrsim 1}=0.05$) we estimate: $\left( \eta _{0}\left( h\right)
x_{0}/h^{2}\varepsilon _{h}\right) _{h\gtrsim 1}\approx 1.3$, so that we
find the expected relation:

\begin{equation}
q_{c}=\left( \frac{\eta _{0}\left( h\right) x_{0}}{\varepsilon _{h}h^{2}}%
\right) _{h\gtrsim 1}^{1/2}\widetilde{\xi }^{-1}\left( t\rightarrow 0\right)
\approx \widetilde{\xi }^{-1}\left( t\rightarrow 0\right)  \label{q_c}
\end{equation}

\section{Quantum tunneling and pair breaking in the boson-insulating state}

It is evident that the ultimately divergent negative conductance implied by
Eq.(\ref{sig^fluctdt0}) is an unphysical result, which clearly indicates the
breakdown of the thermal fluctuations approach at finite field and very low
temperatures. In particular, the unlimited rising Cooper-pairs density
within mesoscopic puddles, predicted by Eq.(\ref{sig^fluctdt0}) in the zero
temperature limit, can be stopped only by pair breaking into unpaired mobile
electron states. Within the fully microscopic (diagrammatic) theory of
fluctuations in superconductors \cite{LV05}, quantum fluctuations associated
with renormalization of the pairing vertices by impurity-scattering (see,
e.g. \cite{GalitLarkinPRB01},\cite{Glatzetal2011}) can lead to such
pair-breaking processes. However, their apparent dynamical nature have not
been treated consistently in the current literature (see, e.g. the
calculation of the DOS contribution in Ref.\cite{LV05}).

Furthermore, the state of the art of the microscopic theory of fluctuations
in superconductors is not sufficiently developed to include dynamical
quantum tunneling of Cooper-pair fluctuations \cite{MZPRB2021}, a phenomenon
which should intensify concurrently with the field-induced pair breaking
processes, due to the strongly enhanced Cooper-pair fluctuations density in
mesoscopic puddles.

Thus, in the absence of a complete microscopic quantum theory of
fluctuations the bosonic TDGL functional approach employed here is
complemented with a phenomenological scheme, which introduces quantum
tunneling of Cooper-pair fluctuations jointly with the dynamical
pair-breaking corrections.

Within this phenomenological approach, we identify in both Eqs.(\ref%
{momdistr}) and (\ref{Q_AL}), "external" and "internal" links for quantum
tunneling corrections to be inserted into both the DOS and the AL
conductivities, respectively. For the DOS conductivity the "external" link
in the momentum distribution function, Eq.(\ref{momdistr}), is the inverse
thermal-prefactor: $1/k_{B}T$, which is interpreted as a characteristic
thermal activation time $\tau _{T}=\hbar /k_{B}T$, whereas the "internal"
link is in the fluctuation energy function $\Phi \left( x;\varepsilon
_{H}\right) \simeq \varepsilon _{H}+\eta \left( H\right) x$. The correction
in the "external" link amounts to modifying $\tau _{T}$ by including the
effect of quantum tunneling through the rate equation:%
\begin{equation}
\frac{1}{\tau _{U}}=\frac{1}{\tau _{T}}+\frac{1}{\tau _{Q}}=k_{B}\left(
T+T_{Q}\right) /\hbar  \label{InvTimeRel}
\end{equation}%
where $\tau _{Q}\equiv \hbar /k_{B}T_{Q}$ is the quantum tunneling time.

The corresponding correction in the "internal" link reflects the dynamics of
the quantum tunneling by shifting the fermionic Matsubara frequency $\omega
_{n}=\left( 2n+1\right) \pi k_{B}T/\hbar $ with the "excitation" frequency $%
\pi k_{B}T_{Q}/\hbar $, under summation defining the digamma functions in
Eqs.(\ref{Phi(x,mu)}) and (\ref{eps_H}). \ 

For the AL conductivity the "external" link in the current correlator Eq.(%
\ref{Q_AL}) is the thermal-rate prefactor for the charge transfer: $%
k_{B}T\propto 1/\tau _{T}$, which is corrected by adding the quantum
tunneling attempt rate $1/\tau _{Q}\propto k_{B}T_{Q}$ according to the rate
equation (\ref{InvTimeRel}), whereas the "internal" links in the fluctuation
energy functions and their derivatives are corrected in a way identical to
that employed for the DOS conductivity (see also Appendix B for more
details).

The over all "external" modifications result in multiplying the AL
conductivity (Eq.(\ref{sig_ALlin})) and dividing the DOS conductivity (Eq.(%
\ref{sig_DOS})) by the same factor $\left( 1+T_{Q}/T\right) $. The
corresponding "internal" modifications, result in shifting the arguments of
the digamma functions and their derivatives in $\varepsilon _{h}$, and $\eta
\left( h\right) $, respectively, with the normalized "excitation" frequency
term $T_{Q}/2T$, which reflect the dynamical nature of the quantum tunneling
introduced to the "external" links. \ 

This pattern of quantum corrections is consistent with the introduction of
the unified quantum-thermal (QT) fluctuations partition function:

\begin{eqnarray}
Z_{fluct} &=&\prod\limits_{\mathbf{q}}\int \mathcal{D}\Delta \left( q\right) 
\mathcal{D}\Delta ^{\ast }\left( q\right)  \label{Zfluct} \\
&&\exp \left\{ -\frac{\tau _{U}}{\hbar }\left[ \widetilde{\varepsilon }%
_{h}^{U}+\frac{\eta _{U}\left( h\right) }{4\pi k_{B}T}Dq^{2}\right]
\left\vert \Delta \left( q\right) \right\vert ^{2}\right\}  \notag
\end{eqnarray}%
where $\tau _{U}$, defined in Eq.(\ref{InvTimeRel}), is interpreted as the
combined QT characteristic time for both activation over and tunneling
through the GL energy barriers separating superconducting and normal state
regimes. The significance of the unified QT electron pairing functions $%
\varepsilon _{h}^{U}$, $\eta _{U}\left( h\right) $, following the "internal"
modifications mentioned above, will be further elaborated below. The
partition function, Eq.(\ref{Zfluct}) yields the QT fluctuations propagator:%
\begin{equation}
D_{U}\left( q;\widetilde{\varepsilon }_{h}^{U}\right) =\frac{k_{B}\left(
T+T_{Q}\right) }{N_{2D}\left( \widetilde{\varepsilon }_{h}^{U}+\frac{%
Dq^{2}\eta _{U}\left( h\right) }{4\pi k_{B}T}\right) }  \label{D_U}
\end{equation}%
in which the "dressed" critical shift parameter, $\widetilde{\varepsilon }%
_{h}^{U}$, due to interaction between Gaussian fluctuations, is determined
from the self-consistent field (SCF) equation \cite{MZPRB2021}: 
\begin{equation}
\widetilde{\varepsilon }_{h}^{U}=\varepsilon _{h}^{U}+\alpha F_{U}\left(
h\right) \left( 1+T_{Q}/T\right) \ln \left( 1+\frac{\eta _{U}\left( h\right)
x_{0}}{\widetilde{\varepsilon }_{h}^{U}t}\right)  \label{SCFeqU}
\end{equation}%
Here:%
\begin{equation}
F_{U}\left( h\right) =\frac{1}{\eta _{U}\left( h\right) }\sum\limits_{n=0}^{%
\infty }\frac{\varkappa _{n}^{U}\left[ \left( \varkappa _{n}^{U}\right) ^{2}+%
\overline{\mu }^{2}h^{2}\right] }{\left[ \varkappa _{n}^{U}\left( \varkappa
_{n}^{U}-2\beta \right) +\overline{\mu }^{2}h^{2}\right] ^{3}},
\label{F_U[h]}
\end{equation}%
with: 
\begin{equation}
\varkappa _{n}^{U}=n+1/2\mathbf{+}T_{Q}/2T+2\beta +\overline{\delta }h^{2}
\label{chi_n^U}
\end{equation}%
is the four-electron correlator controlling the interaction between
fluctuations, and:

\begin{equation}
\alpha \equiv \frac{1}{\hslash \pi ^{3}DN_{2D}}=\frac{2}{\pi ^{2}}\left( 
\frac{\varepsilon _{SO}}{E_{F}}\right)  \label{alpha}
\end{equation}%
is the interaction strength parameter. \ Note the "external" quantum
tunneling correction factor $\left( 1+T_{Q}/T\right) $ in Eq.(\ref{SCFeqU}),
which originates in the unified QT rate factor of the fluctuation
propagator, as written in Eq.(\ref{D_U}). The "external" corrections to the
AL and the DOS conductivities in Eq.(\ref{sig_ALlin}) and Eq.(\ref{sig_DOS})
respectively are equivalent to replacing the stiffness parameter appearing
in their prefactors with the hybrid expression: 
\begin{equation}
\eta \left( h\right) \rightarrow \left( 1+\frac{T_{Q}}{T}\right) \eta
_{U}\left( h\right)  \label{eta_correct}
\end{equation}%
where:

\begin{equation}
\eta _{U}\left( h\right) =\sum\limits_{n=0}^{\infty }\frac{\left( \varkappa
_{n}^{U}\right) ^{2}-\overline{\mu }^{2}h^{2}}{\left[ \varkappa
_{n}^{U}\left( \varkappa _{n}^{U}-2\beta \right) +\overline{\mu }^{2}h^{2}%
\right] ^{2}}  \label{eta_U[h]}
\end{equation}

The "excitation" frequency-shift term $T_{Q}/2T$ appearing in Eq.(\ref%
{eta_U[h]}) (through Eq.(\ref{chi_n^U})), represents pair-breaking effect
associated with the tunneling process. This is seen more directly under the
transformation $\varepsilon _{H}\rightarrow \varepsilon _{h}^{U}$ of the
critical-shift parameter: 
\begin{eqnarray}
\varepsilon _{h} &\rightarrow &\varepsilon _{h}^{U}\equiv \ln \left( \frac{T%
}{T_{c0}}\right) +a_{+}\psi \left( \frac{1}{2}+T_{Q}/2T+f_{-}\right)  \notag
\\
&&+a_{-}\psi \left( \frac{1}{2}+T_{Q}/2T+f_{+}\right) -\psi \left( 1/2\right)
\label{eps_Hcorr}
\end{eqnarray}

In the absence of quantum tunneling $\varepsilon _{h}$ (Eq.(\ref{eps_H})) is
subjected to the usual magnetic field induced pair-breaking effect \cite%
{ShahLopatin07} through the Zeeman spin-splitting energy ($\mu _{B}H$)\ and
the diamagnetic energy ($D\left( deH\right) ^{2}/\hslash $) terms. In the
zero temperature limit, the effect is dramatically reflected in the removal
of the (Cooper) singularity of the logarithmic term in Eq.(\ref{eps_H}), due
to exact cancellation by the asymptotic values of the digamma functions for $%
\ f_{\pm }\gg 1$ (see Appendix C). In the presence of quantum tunneling, the
excitation frequency shift $\pi k_{B}$$T_{Q}/\hbar $\ introduced to define $%
\varepsilon _{h}^{U}$, Eq.(\ref{eps_Hcorr}), causes in this limit an
additional, pair-breaking effect, not driven directly by magnetic field,
through the asymptotic behavior of the digamma functions for $T_{Q}/2T\gg 1$%
\ (see Appendix C).

For systems with long range phase coherence described, e.g. in Ref.\cite%
{ShahLopatin07}, \cite{Lopatinetal05} the main impact of the pair-breaking
perturbations is near the critical point $\varepsilon _{h}=0$ for Cooper
pairs condensation (at $q=0)$ in momentum space. For the boson system of
strong SC fluctuations at very low temperatures, under consideration here,
the softening of the fluctuation modes, that follows the critical pair
breaking near $q=0$, takes place over a large range of wavenumbers, where
Cooper pairs tend to condense within mesoscopic puddles in real space. The
excitation processes represented by the Matsubara frequency shift $\pi
k_{B}T_{Q}/\hbar $, associated with the dynamical quantum tunneling
processes represented by the $1+T_{Q}/T$ factor, yield partial recovery of
the stiffness parameter at finite fields (see Fig.1), and so suppress the
Cooper-pair fluctuations density and reinforce pair-breaking into unpaired
electron states.

To summarize, the frequency shift that transforms $\eta \left( h\right) $ to 
$\eta _{U}\left( h\right) $ and represents pair breaking effect, is
intimately connected to the quantum tunneling process discussed above. This
is clearly seen by considering the zero temperature limit of $\eta
_{U}\left( h\right) $ in Eq.(\ref{eta_U[h]}):

\begin{equation}
\left( \eta _{U}\left( h\right) \right) _{T\rightarrow 0}=\left( \frac{T}{%
T_{Q}}\right) _{T\rightarrow 0}\eta _{Q}\left( h\right)  \label{QLeta_U}
\end{equation}%
where: {\small 
\begin{eqnarray}
&&\eta _{Q}\left( h\right) \equiv \int\limits_{0}^{\infty }d\nu \frac{\left(
\varkappa _{\nu }^{Q}\right) ^{2}-\mu _{Q}^{2}h^{2}}{\left[ \varkappa _{\nu
}^{Q}\left( \varkappa _{\nu }^{Q}-2\beta _{Q}\right) +\mu _{Q}^{2}h^{2}%
\right] ^{2}}  \label{eta_Q(h)} \\
&&=\frac{1/2+2\beta _{Q}+\delta _{Q}h^{2}}{\left( 1/2+\delta
_{Q}h^{2}\right) \left( 1/2+\delta _{Q}h^{2}+2\beta _{Q}\right) +\mu
_{Q}^{2}h^{2}}  \notag
\end{eqnarray}%
}and $\varkappa _{\nu }^{Q}=\nu +1/2+2\beta _{Q}+\delta _{Q}h^{2}$, with $%
\beta _{Q}=\beta _{0}/t_{Q},\mu _{Q}=\mu _{0}/t_{Q},\delta _{Q}=\delta
_{0}/t_{Q},t_{Q}\equiv T_{Q}/T_{c}^{\ast }$.

The limiting function $\eta _{Q}\left( h\right) $ in Eq.(\ref{eta_Q(h)}) is
a continuous smooth function of the field $h$, including at $h=0$. \
Therefore, Eq.(\ref{QLeta_U}) implies that the discontinuous plunge of $\eta
\left( h\right) $ at $h=0$ in the zero temperature limit (see Fig.1) is
removed by the frequency shift term, as can be directly checked in Eq.(\ref%
{eta_U[h]}). The magnitude of $\eta _{U}\left( h\right) $ diminishes
uniformly to zero with $T/T_{Q}$ in this limit. However, by multiplying with
the divergent quantum tunneling factor $\left( 1+T_{Q}/T\right) $ the
resulting hybrid product in Eq.(\ref{eta_correct}), which represents the
combined effect of quantum tunneling and pair breaking, is a smooth finite
function of the field $\eta _{Q}\left( h\right) $ (see Appendix C).

\begin{figure}[tbh]
\includegraphics[width =.45\textwidth]{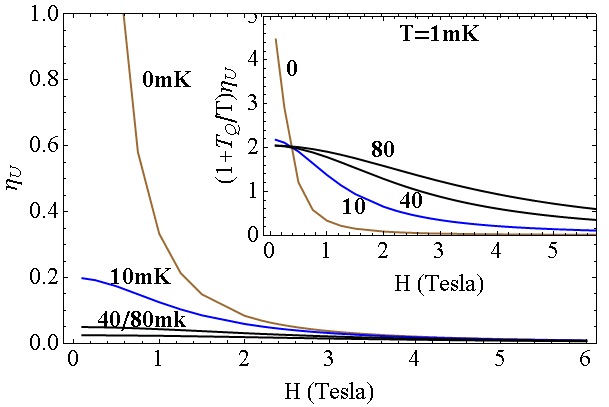}
\caption{Field-dependent stiffness parameter $\protect\eta _{U}\left(
H\right) $ calculated at $T=1$ mK for $T_{Q}=0,10,40,80$ mK. Inset: The
hybrid product $\left( 1+T_{Q}/T\right) \protect\eta _{U}\left( H\right) $
calculated for the same $T$ and $T_{Q}$ values as presented in the main
figure.}
\end{figure}

A similar hybrid form and limiting behavior at zero temperature characterize
the interaction term in the SCF equation (\ref{SCFeqU}). The four-electron
correlator: \ 
\begin{equation}
\left( F_{U}\left( h\right) \right) _{T\rightarrow 0}=\left( \frac{T}{T_{Q}}%
\right) _{T\rightarrow 0}F_{Q}\left( h\right)  \label{QLF_U}
\end{equation}%
where: 
\begin{equation}
F_{Q}\left( h\right) \equiv \frac{1}{\eta _{Q}\left( h\right) }%
\int\limits_{0}^{\infty }d\nu \frac{\varkappa _{\nu }^{Q}\left[ \left(
\varkappa _{\nu }^{Q}\right) ^{2}+\mu _{Q}^{2}h^{2}\right] }{\left[
\varkappa _{\nu }^{Q}\left( \varkappa _{\nu }^{Q}-2\beta _{Q}\right) +\mu
_{Q}^{2}h^{2}\right] ^{3}}  \label{F_Q(h)}
\end{equation}%
shows similar singular behavior to that of $\eta _{U}\left( h\right) $ (see
Eqs.(\ref{QLeta_U}) and (\ref{eta_Q(h)})). The overall interaction term has
the finite regular limiting form (including the logarithm):

\begin{eqnarray}
&&\alpha F_{U}\left( h\right) \left( 1+T_{Q}/T\right) \ln \left( 1+\frac{%
\eta _{U}\left( h\right) x_{0}t^{-1}}{\widetilde{\varepsilon }_{h}^{U}}%
\right)  \notag \\
&&\rightarrow \alpha F_{Q}\left( h\right) \ln \left( 1+\frac{\eta _{Q}\left(
h\right) x_{0}}{\widetilde{\varepsilon }_{h}^{Q}t_{Q}}\right)  \label{QLint}
\end{eqnarray}

This SCF approach avoids the critical divergence of both the AL
paraconductivity and the DOS conductivity, and allows to extend the
expression for the conductance fluctuations $\sigma ^{fluct}d=\sigma
_{AL}d+\delta \sigma _{DOS}d$, given in terms of Eqs.(\ref{sig_DOS}),(\ref%
{sig_ALlin}), to regions well below the nominal critical SC transition. It
also offers an extended proper measure of the{\normalsize \ }pair-breaking
effect. In contrast to $\varepsilon $$_{h}$, $\widetilde{{\normalsize %
\varepsilon }}_{h}$ is positive definite in the entire fields range,
including that below the critical field where $\varepsilon $$_{h}<0$ (see
Fig.2). The uniform enhancement of $\widetilde{{\normalsize \varepsilon }}$$%
_{h}^{U}$ with respect to $\widetilde{{\normalsize \varepsilon }}_{h}$, seen
in Fig.2, resulting from the introduction of the frequency shift to the SCF
equation (\ref{SCFeqU}), is a genuine measure of the pair-breaking effect
associated with the quantum tunneling. Its monotonically increasing field
dependence seen in Fig.2 properly reflects the field-induced pair-breaking
effect in the entire fields range.

\begin{figure}[tbh]
\includegraphics[width =.45\textwidth]{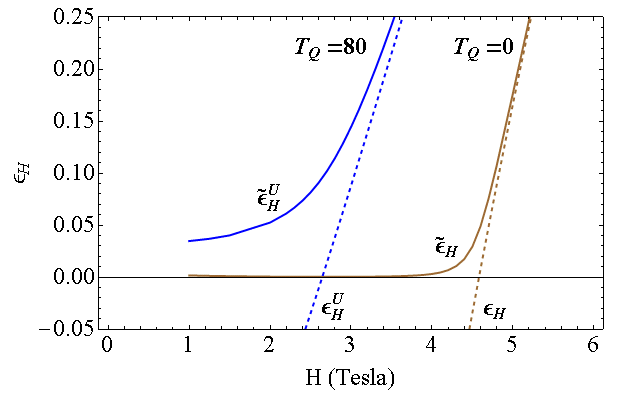}
\caption{Field dependence, at $T=30$ mK, of the "bare" critical-shift
parameter $\protect\varepsilon _{H}^{U}$ (dashed lines), and the
corresponding self-consistently "dressed" parameter $\widetilde{\protect%
\varepsilon }_{H}^{U}$ (solid lines), in the absence of quantum tunneling
(brown curves) and for $T_{Q}=80$ mK (blue curves). \ Note the downward
shift of the critical field and the uniform enhancement of the dressed
critical-shift parameter associated with the quantum tunneling effect.}
\end{figure}

\section{Discussion}

In this paper we have discovered, while searching for the deep origin of the
high-field insulating states appearing at diminishing temperature, that due
to extreme softening of the fluctuation modes and their redistribution over
a large region in momentum space, there is a propensity of Cooper-pair
fluctuations to condense in real-space puddles of decreasing spatial size, $%
\widetilde{\xi }\left( t\rightarrow 0\right) $ (see Eq.(\ref{xit0})). This
picture is of course ideal, but basically reflects real tendency toward
boson insulating states. Charge transfers between the exclusive bosons
system and the normal-electron states, underlying the microscopic Gorkov GL
approach employed, are introduced within a unified phenomenological
approach, by allowing field-induced pair-breaking processes to develop
during dynamical quantum tunneling of Cooper-pair fluctuations out of the
mesoscopic puddles.

Other quantum fluctuations effects arising from coherent Andreev-like
scattarings \cite{Lopatinetal05}, \cite{LV05}, \cite{ManivAlexander76},
associated with the Maki-Thompson (MT) contribution to the paraconductivity 
\cite{MakiPTP1968}, \cite{ThompsonPRB1970}, are expected to be suppressed by
strong spin-orbit scatterings \cite{LV05}, which characterize the SrTiO$_{3}$%
/LaAlO$_{3}$ interfaces under consideration here \cite{Mograbi19}, \cite%
{RoutPRL2017}.

Exploiting the complete agreement between the results of our approach and
those of the fully microscopic theory at zero magnetic field, it will be
meaningful at this point to compare the influence of the quantum
fluctuations employed in each approach on the conductivity at finite field.
Thus, on one hand, the DOS conductivity derived in our approach in the
quantum limit (see Eq.\ref{sig_DOS^Q}), and the renormalized single-particle
conductivity derived within the fully microscopic approach in the quantum
fluctuations regime \cite{Glatzetal2011}, are both finite, with negative
sign, and have the same field dependence. On the other hand, in the fully
microscopic approach the vanishing rate of the AL paraconductivity is
further accelerated in the quantum fluctuations regime \cite{Glatzetal2011},
whereas in our approach the vanishing AL conductivity (see Eq.(\ref%
{sig^fluctdt0})) is recovered by the effect of quantum tunneling (see Eq.\ref%
{sig_AL^Q}). The physical reasoning behind this recovery is explained in
Sec.IV and in Appendix B.

An important feature of the localization process predicted in our approach
is its dynamical nature, namely that it{\normalsize \ }occurs in response to
the driving electric force \cite{MZPRB2021}, and not spontaneously in a
thermodynamical process toward equilibrium state. This feature seems to
distinguish it from the various approaches to the phenomenon of SIT
discussed in the literature \cite{Dubi07}, \cite{Bouadim2011}, \cite%
{GhosalPRL98}, \cite{Vinokur2008}, in which disorder-induced spatial
inhomogeneity in the form of SC islands is involved in generating the
insulating state. However, in a similar manner the formation of fluctuation
puddles in our approach is controlled by disorder, which strongly affect the
Cooper-pairs amplitude correlation function in real space. This can be seen
by comparing the pair correlation function derived in the dirty limit \cite%
{MZPRB2021},\cite{CaroliMaki67I} to that obtained in the pure limit \cite%
{CaroliMaki67II}.

Another important parameter in our approach of relevance to the insulating
behavior that seems to have a parallel in the literature \cite{Bouadim2011},
is the self-consistent critical shift parameter $\widetilde{\varepsilon }_{H}
$, which also plays the role of an energy gap in the Cooper-pair
fluctuations spectrum \cite{MZPRB2021}. Thus, it is interesting to note that
the two-particle gap, which characterizes the insulating state in Ref.\cite%
{Bouadim2011}, vanishes at the SIT. Analogously, in our approach the
(two-particle) Cooper-pair fluctuation gap $\widetilde{\varepsilon }_{H}$
gradually diminishes to very small (nonvanishing)\ values upon decreasing
field below the sheet-resistance peak (see Fig.2), in accord with the lack
of a critical point.

The combined effect of this nonvanishing two-particle gap $\widetilde{%
\varepsilon }_{H}$ and the diminishing stiffness parameter $\eta \left(
H\right) $ upon increasing field at very low temperatures, is responsible
for the loss of long-range phase coherence and for the puddles formation.
The resulting boson insulating state is reminiscent of the field-induced
paired insulating phase discussed in Ref.\cite{Datta-arXiv2021}, which is
also closely related to the picture of the suppressed Bose insulator
deliberated in Ref.\cite{BaturinaPRL2007}. The introduction of quantum
tunneling of Cooper-pair fluctuations within our complementary
phenomenological approach, which leads to the broadening of the sharp MR
peaks at low temperatures, is clearly consistent with the conducting
Josephson tunneling effect among SC islands.

Finally, a few comments about the robustness of the quantitative comparison
with the experimental data \cite{Mograbi19} are in order. As the direct
analytic continuation scheme of the AL correlator employed in Ref.\cite%
{MZPRB2021} (see the remarks below Eq.(\ref{sig_ALlin})) doubles the
prefactor of the corresponding paraconductivity as compared to the
well-known result, the implication for the fitting process of using the
latter prefactor is expected to further amplify the relative magnitude of
the negative DOS conductivity and so to further reinforce the appearance of
the field-induced boson insulating states at low temperatures. This has been
confirmed quantitatively in Appendix D by repeating the fitting process
described in detail in Ref.\cite{MZPRB2021} with the 1/2 prefactor of the AL
contribution. The results show that good agreement with the experimental
data can be achieved also with the reduced AL prefactor by changing the
spin-orbit scattering parameter moderately within its range of uncertainty.

\bigskip

\section{Acknowledgments}

We would like to thank Eran Maniv, Itai Silber and Yoram Dagan for useful
discussions. We are also indebted to Andrei Varlamov for helpful comments.

\appendix

\section{Range of validity of the linear approximation}

Extending the first order expansion to next order, i.e. writing: $\Phi
\left( x;\varepsilon _{H}\right) =\varepsilon _{H}+\eta \left( H\right)
x+\zeta \left( H\right) x^{2}+...$, where $\zeta \left( H\right) =\left[
a_{+}\psi ^{\prime \prime }\left( 1/2+f_{-}\right) +a_{-}\psi ^{\prime
\prime }\left( 1/2+f_{+}\right) \right] /2$, we find at low temperature ($%
t\ll 1$) and finite magnetic field, $h>0$, that in addition to Eq.(\ref%
{eta(h)t0}), Eq.(\ref{eta_0(h)}), $\zeta \left( H\right) \rightarrow \zeta
_{0}\left( h\right) t^{2}/h^{4}$, where:

{\small 
\begin{eqnarray*}
&&\frac{1}{h^{4}}\zeta _{0}\left( h\right) =\frac{1}{2} \left[
\int\limits_{0}^{\infty }d\nu \frac{\left( 1+\beta _{0}/\sqrt{\beta
_{0}^{2}-\mu _{0}^{2}h^{2}}\right)}{\left( \nu +\delta _{0}h^{2}+\beta _{0}-%
\sqrt{\beta _{0}^{2}-\mu _{0}^{2}h^{2}}\right) ^{3}}\right. \\
&&\left.+ \int\limits_{0}^{\infty }d\nu \frac{\left( 1-\beta _{0}/\sqrt{%
\beta _{0}^{2}-\mu _{0}^{2}h^{2}}\right)}{\left( \nu +\delta _{0}h^{2}+\beta
_{0}+\sqrt{\beta _{0}^{2}-\mu _{0}^{2}h^{2}}\right) ^{3}}\right]
\end{eqnarray*}
}

\bigskip

Evaluation of the integral leads to:

\begin{equation*}
\zeta _{0}\left( h\right) =\frac{1}{2}\frac{\left( 2\beta _{0}+\delta
_{0}h^{2}\right) ^{2}-\mu _{0}^{2}h^{2}}{\left[ \left( 2\beta _{0}+\delta
_{0}h^{2}\right) \delta _{0}+\mu _{0}^{2}\right] ^{2}}
\end{equation*}%
so that the relevant expansion at high fields ($h\sim 1$), is: 
\begin{equation}
\Phi \left( x;\varepsilon _{h}\right) \rightarrow \varepsilon _{H}+\frac{1}{%
h^{2}}\eta _{0}\left( h\right) \left( tx\right) +\frac{1}{h^{4}}\zeta
_{0}\left( h\right) \left( tx\right) ^{2}+...  \label{PhiExpt0}
\end{equation}%
with the corresponding temperature independent expansion parameter:

\begin{equation*}
xt=\frac{\hbar Dq^{2}}{4\pi k_{B}T_{c}^{\ast }}\leq \frac{\hbar Dq_{c}^{2}}{%
4\pi k_{B}T_{c}^{\ast }}=x_{0}
\end{equation*}

For the experimental situation encountered in Ref.\cite{MZPRB2021} the
diamagnetic energy term $\delta _{0}h^{2}$ is much smaller than both the
spin-orbit energy $\beta _{0}$ and the Zeeman splitting $\mu _{0}h$,
implying that the coefficients: $\eta _{0}\left( h\right) \simeq 2\beta
_{0}/\mu _{0}^{2},\zeta _{0}\left( h\right) \simeq \left( 4\beta
_{0}^{2}-\mu _{0}^{2}h^{2}\right) /2\mu _{0}^{4}$, are constant, or nearly
constant (since typically $\left( 2\beta _{0}\right) ^{2}\gg \left( \mu
_{0}h\right) ^{2}$).

We therefore conclude that the condition for uniform convergence of the
expansion is $x_{0}\ll 1$. \ In our fitting process we have used the value $%
x_{0}=0.015$, well within the domain of convergence.

\bigskip

\section{The quantum fluctuations corrections to conductivity}

In this appendix we outline the physical reasoning behind our
phenomenological quantum fluctuations correction to the two ingredients of
the conductance fluctuations. Starting with the DOS conductivity we consider
the Cooper-pair density, $n_{s}$, given in Eq.(\ref{n_s}), with $%
\left\langle \left\vert \phi \left( q\right) \right\vert ^{2}\right\rangle $
in Eq.(\ref{momdistr}). Approximating $7\zeta \left( 3\right) \simeq 8.4$ we
rewrite:

\begin{equation}
\left\langle \left\vert \phi \left( q\right) \right\vert ^{2}\right\rangle
\simeq 4.2\frac{\left( n_{2D}\lambda _{T}^{2}\right) }{\Phi \left(
x;\varepsilon _{H}\right) }  \label{MomDist}
\end{equation}%
where \ $\Phi \left( x;\varepsilon _{H}\right) \simeq \varepsilon _{H}+\eta
\left( H\right) x$,\ $n_{2D}=k_{F}^{2}/2\pi $ is the density of the 2D
electron gas and $\lambda _{T}=\sqrt{\hbar ^{2}/2\pi m^{\ast }k_{B}T}$ is
the thermal wavelength.

The momentum distribution function $\left\langle \left\vert \phi \left(
q\right) \right\vert ^{2}\right\rangle $ measures the number of bosons per
wave vector $\mathbf{q}$ in the Cooper-pairs liquid, engaged in equilibrium
with a 2D gas of unpaired mobile electrons with a nominal density $n_{2D}$.
The prefactor $n_{2D}\lambda _{T}^{2}=\left( 1/2\pi ^{2}\right) \left(
E_{F}\tau _{T}/\hbar \right) $, that is the number of electrons in an area
of size equal to the thermal wavelength, is proportional to the
characteristic thermal activation time $\tau _{T}=\hbar /k_{B}T$.

The quantum corrections, introduced in the main text, amount to modifying
Expression (\ref{MomDist}) in two steps; in the first, replacing the
temperature $T$, appearing in the denominator of the prefactor, with $%
T+T_{Q} $, and in the second step inserting the effective frequency-shift
term $T_{Q}/2T$\ to the arguments of the digamma functions in Eq.(\ref%
{Phi(x,mu)}) consistently with the replacement of $\varepsilon _{H}$ with $%
\varepsilon _{H}^{U}$. The total modification takes the form:

\begin{eqnarray*}
\left\langle \left\vert \phi \left( q\right) \right\vert ^{2}\right\rangle
&\rightarrow &\left\langle \left\vert \phi _{U}\left( q\right) \right\vert
^{2}\right\rangle =n_{2D}\lambda _{U}^{2}\frac{4.2}{\Phi _{U}\left(
x;\varepsilon _{H}^{U}\right) } \\
&=&\frac{2.1}{\pi ^{2}}\frac{\left( E_{F}\tau _{U}/\hbar \right) }{\Phi
_{U}\left( x;\varepsilon _{H}^{U}\right) }
\end{eqnarray*}%
where $\Phi _{U}\left( x;\varepsilon _{H}^{U}\right) \simeq \varepsilon
_{H}^{U}+\eta _{U}\left( H\right) x$. The prefactor $n_{2D}\lambda _{U}^{2}$%
, is the number of electrons in an effective area $\lambda _{U}^{2}=\hbar
^{2}/2\pi m^{\ast }k_{B}\left( T+T_{Q}\right) $ that is proportional to the
characteristic time, $\tau _{U}$, for both thermal activation and quantum
tunneling of Cooper pairs. Thus, increasing the temperature and/or
shortening the time $\tau _{Q}$ for quantum tunneling (which also enhance
pair breaking by increasing $\Phi _{U}\left( x;\varepsilon _{H}^{U}\right) $%
), result in larger rate of thermal and/or quantum leakage from puddles of
Cooper pairs. The resulting reduction in the number of Cooper-pairs, which
occurs versus a corresponding increase in the number of unpaired mobile
electrons, would suppress the DOS contribution to the resistance.

The corresponding unified (quantum-thermal (QT)) density (per unit area) of
the Cooper-pairs liquid is now evaluated: $n_{s}^{U}=\frac{1}{d}\frac{1}{%
\left( 2\pi \right) ^{2}}\int \left\langle \left\vert \phi _{U}\left(
q\right) \right\vert ^{2}\right\rangle d^{2}q=\frac{1}{d}\frac{1}{\left(
2\pi \right) ^{2}}\int_{0}^{q_{c}^{2}}\pi d\left( q^{2}\right) \left( \frac{%
2.1E_{F}}{\pi ^{2}k_{B}\left( T+T_{Q}\right) }\right) \frac{1}{\Phi
_{U}\left( x;\varepsilon _{H}^{U}\right) }$, so that the unified DOS
conductivity, $\sigma _{DOS}^{U}=-\left( 2n_{s}^{U}e^{2}/m^{\ast }\right)
\tau _{SO}$, is given by:

\begin{equation}
\sigma _{DOS}^{U}d\simeq -4.2\left( \frac{G_{0}}{\pi }\right)
\int_{0}^{t^{-1}x_{0}}\frac{dx}{\left( 1+T_{Q}/T\right) \Phi _{U}\left(
x;\varepsilon _{H}^{U}\right) }  \label{sig_DOS^U}
\end{equation}

For the AL thermal fluctuations conductivity we start with the retarded
current-current correlator $Q_{AL}^{R}\left( \Omega \right) $, Eq.(\ref{Q_AL}%
), which was obtained from the Matsubara correlator $Q_{AL}\left( i\Omega
_{\nu }\right) $ following the analytic continuation $i\Omega _{\nu
}\rightarrow \Omega $. \ The corresponding electrical response function is
seen to be proportional to the thermal energy $k_{B}T$. \ The effects of
quantum tunneling and pair breaking are introduced by adding to the thermal
attempt rate $1/\tau _{T}\propto k_{B}T$ the quantum tunneling attempt rate $%
1/\tau _{Q}\propto k_{B}T_{Q}$ , and by appropriately inserting the
effective frequency-shift term $T_{Q}/2T$ \ into the function $\Phi \left(
x+\left\vert \mu +\nu \right\vert ;\varepsilon _{H}\right) $, as explained
in the main text, i.e.:


\begin{eqnarray*}
&&Q_{AL}^{U}\left( i\Omega _{\nu }\right) =k_{B}\left( T+T_{Q}\right) \left( 
\frac{2e}{\hbar }\right) ^{2}\left( \frac{1}{2\pi d}\right)\int%
\limits_{0}^{x_{c}}xdx \times \\
&&\sum\limits_{\mu =0,\pm 1,\pm 2,....}\frac{\Phi _{U}^{\prime }\left(
x+\left\vert \mu +\nu \right\vert ;\varepsilon _{H}^{U}\right) }{\Phi
_{U}\left( x+\left\vert \mu +\nu \right\vert ;\varepsilon _{H}^{U}\right) }%
\frac{\Phi _{U}^{\prime }\left( x+\left\vert \mu \right\vert ;\varepsilon
_{H}^{U}\right) }{\Phi _{U}\left( x+\left\vert \mu \right\vert ;\varepsilon
_{H}^{U}\right) }
\end{eqnarray*}

Now, by repeating the procedure employed in deriving Eq.(\ref{sig_ALd}), in
which the above discrete summation is directly continued analytically, $\nu
\rightarrow \hbar \Omega /2\pi ik_{B}T$, and expanded about zero frequency,
we arrive at the following expression for the unified QT AL conductivity:

\begin{equation}
\sigma _{AL}^{U}d=\frac{1}{4}\left( \frac{G_{0}}{\pi }\right) \left( 1+\frac{%
T_{Q}}{T}\right) \int\limits_{0}^{t^{-1}x_{0}}\left( \frac{\Phi _{U}^{\prime
}\left( x;\varepsilon _{H}^{U}\right) }{\Phi _{U}\left( x;\varepsilon
_{H}^{U}\right) }\right) ^{2}dx  \label{sig_AL^U}
\end{equation}

As indicated in Sec.IIIB below Eq.(\ref{sig_ALlin}), the common scheme of
analytic continuation utilizing the contour integration method for
performing the Matsubara summation yields the same result as Eq.(\ref%
{sig_AL^U}) but with the prefactor 1/4 replaced with 1/8.

\bigskip


\section{The quantum limit}

In this appendix we examine the zero-temperature (quantum) limit of the
conductance fluctuation analyzed in Appendix B. We begin by studying the
field-induced pair-breaking and quantum tunneling of Cooper pairs in this
limiting situation. \ Consider the unified (quantum-thermal) expression, Eq.(%
\ref{eps_Hcorr}), for the critical shift parameter $\varepsilon _{h}^{U}$.
\bigskip\ Using the asymptotic expansion of $\psi \left( \frac{1}{2}%
+T_{Q}/2T+f_{\pm }\right) $ for $T_{Q}/T,$ $f_{\pm }\gg 1$, i.e. $\psi
\left( \frac{1}{2}+T_{Q}/2T+f_{\pm }\right) \rightarrow \ln \left(
T_{Q}/2T+f_{\pm }\right) =\ln \left[ \left( T_{Q}+T_{\pm }\right) /2T\right] 
$, we have: \ 
\begin{eqnarray}
&&\varepsilon _{h}^{U}\rightarrow \varepsilon _{h}^{Q}=\ln \left(
T/T_{c0}\right) -\ln T+  \label{eps_h^UT0} \\
&&a_{+}\ln \left( T_{Q}+T_{-}\right) +a_{-}\ln \left( T_{Q}+T_{+}\right)
-\ln 2-\psi \left( 1/2\right)  \notag
\end{eqnarray}%
where:

\begin{equation}
T_{\pm }\equiv \frac{D\left( de\right) ^{2}H^{2}}{\pi k_{B}\hslash }+\frac{%
\varepsilon _{SO}}{2\pi k_{B}}\pm \sqrt{\left( \frac{\varepsilon _{SO}}{2\pi
k_{B}}\right) ^{2}-\left( \frac{\mu _{B}H}{\pi k_{B}}\right) ^{2}}
\label{T_PM}
\end{equation}

In the above expression for $\varepsilon _{h}^{Q}$ (Eq.(\ref{eps_h^UT0})),
the Cooper singular term, $\ln \left( T/T_{c0}\right) $, is exactly
cancelled by the logarithmic term arising from the asymptotic expansion of
the digamma functions, so that the remaining regular terms are rearranged to
yield the following temperature independent expression for ${\normalsize %
\varepsilon _{h}^{Q}}$: 
\begin{equation}
{\normalsize \varepsilon _{h}^{Q}\rightarrow a_{+}\ln \left( \frac{%
T_{Q}+T_{-}}{T_{c0}}\right) +a_{-}\ln \left( \frac{T_{Q}+T_{+}}{T_{c0}}%
\right) +\ln 2+\gamma }  \label{epas_h^Q}
\end{equation}%
where $\gamma \approx 0.5772$... is the Euler--Mascheroni constant, and:

\begin{equation*}
a_{\pm }=\frac{1}{2}\left( 1\pm \frac{1}{\sqrt{1-\left( \mu _{0}/\beta
_{0}\right) ^{2}h^{2}}}\right)
\end{equation*}

We now turn to the quantum limit of the DOS conductivity, Eq.(\ref{sig_DOS^U}%
): {\small 
\begin{eqnarray*}
&&\int_{0}^{t^{-1}x_{0}}\frac{dx}{\left( 1+T_{Q}/T\right) \Phi _{U}\left(
x;\varepsilon _{H}^{U}\right) }\approx \\
&&\int_{0}^{t^{-1}x_{0}}\frac{dx}{\left( 1+T_{Q}/T\right) \left[ \varepsilon
_{h}^{U}+\eta _{U}\left( h\right) x\right] }= \\
&&\frac{1}{\eta _{U}\left( h\right) \left( 1+T_{Q}/T\right) }\ln \left( 1+%
\frac{\eta _{U}\left( h\right) t^{-1}x_{0}}{\varepsilon _{h}^{U}}\right) ,
\end{eqnarray*}
} so that by going to the limit $T_{Q}/T\rightarrow \infty $, and using Eq.(%
\ref{eta_Q(h)}), i.e.: $\eta _{Q}\left( h\right) =\frac{1/2+2\beta
_{Q}+\delta _{Q}h^{2}}{\left( 1/2+\delta _{Q}h^{2}\right) \left( 1/2+\delta
_{Q}h^{2}+2\beta _{Q}\right) +\mu _{Q}^{2}h^{2}}\approx \frac{2\beta _{Q}}{%
\beta _{Q}+\mu _{Q}^{2}h^{2}}=\frac{t_{Q}\eta _{0}}{t_{Q}\eta _{0}/2+h^{2}}$%
, \ we arrive at the final temperature-independent expression:

\begin{eqnarray}
\sigma _{DOS}^{Q}d &\simeq &-4.2\left( \frac{G_{0}}{\pi }\right) \frac{%
t_{Q}\eta _{0}/2+h^{2}}{t_{Q}\eta _{0}}\times  \notag \\
&&\ln \left[ 1+\frac{\eta _{0}x_{0}}{\widetilde{\varepsilon }_{h}^{Q}\left(
t_{Q}\eta _{0}/2+h^{2}\right) }\right] ,  \label{sig_DOS^Q} \\
\eta _{0}\left( h\right) &\approx &\eta _{0}\equiv \frac{2\beta _{0}}{\mu
_{0}^{2}},G_{0}=\frac{e^{2}}{\pi \hbar }  \notag
\end{eqnarray}%
where $\widetilde{\varepsilon }_{h}^{Q}$ is determined by the SCF equation,
Eq.(\ref{SCFeqU}), in the quantum limit, i.e.:

\begin{equation}
\widetilde{\varepsilon }_{h}^{Q}=\varepsilon _{h}^{Q}+\alpha F_{Q}\left(
h\right) \ln \left[ 1+\frac{\eta _{0}x_{0}}{\widetilde{\varepsilon }%
_{h}^{Q}\left( t_{Q}\eta _{0}/2+h^{2}\right) }\right]  \label{SCFeq_Q}
\end{equation}

\begin{figure}[t]
\includegraphics[width =.45\textwidth]{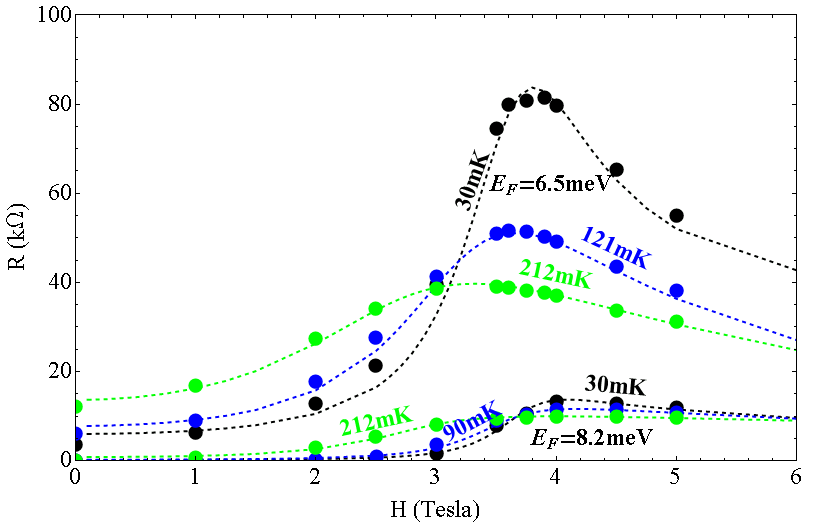}
\caption{Measured sheet resistance as a function of field at different
temperatures for two gate voltages, corresponding to $R_{N}$ $=$ $20.5$ k$%
\Omega $ ($E_{F}$ $=$ $6.5$ meV) and $7.5$ k$\Omega $ ($E_{F}$ $=$ $8.2$
meV) as reported in Ref.\protect\cite{Mograbi19}\ (full circles). The dashed
lines (with temperature labels) represent the results of calculations
similar to those performed in Ref.\protect\cite{MZPRB2021} , but with 1/2 of
the total amplitude of the AL conductivity term used in Ref.\protect\cite%
{MZPRB2021} (see more details in the text of Appendix D). }
\end{figure}

In the absence of the self-consistent interaction between fluctuations the
critical shift parameter reduces to ${\normalsize \varepsilon _{h}^{Q}}$,
which vanishes at the quantum critical field ${\normalsize h}_{c}$ , so that
near ${\normalsize h}_{c}$:

\begin{equation*}
{\normalsize \varepsilon _{h}^{Q}\propto h-h}_{c}{\normalsize \rightarrow 0}
\end{equation*}

It is instructive to note that Eq.\ref{sig_DOS^Q} is equivalent to the
fluctuation conductivity derived in Ref.\cite{Glatzetal2011} in the region
of quantum fluctuations within a fully microscopic (diagrammatic) approach.

Considering the unified AL conductivity, Eq.(\ref{sig_AL^U}) in the linear
approximation:

$\sigma _{AL}^{U}d=\frac{1}{4}\left( \frac{G_{0}}{\pi }\right) \left( 1+%
\frac{T_{Q}}{T}\right) \int\limits_{0}^{t^{-1}x_{0}}\left( \frac{\eta
_{U}\left( h\right) }{\varepsilon _{h}^{U}+\eta _{U}\left( h\right) x}%
\right) ^{2}dx=\frac{1}{4}\left( \frac{G_{0}}{\pi }\right) \left( 1+\frac{%
T_{Q}}{T}\right) \eta _{U}^{2}\left( h\right) \frac{t^{-1}x_{0}}{\varepsilon
_{h}^{U}\left[ \varepsilon _{h}^{U}+\eta _{U}\left( h\right) t^{-1}x_{0}%
\right] }$, so that in the limit, $T_{Q}/T\rightarrow \infty $:

\begin{equation}
\sigma _{AL}^{Q}d\simeq \frac{1}{4}\left( \frac{G_{0}}{\pi }\right) \frac{%
t_{Q}\eta _{0}^{2}x_{0}}{\left( t_{Q}\eta _{0}/2+h^{2}\right) \widetilde{%
\varepsilon }_{h}^{Q}\left[ \left( t_{Q}\eta _{0}/2+h^{2}\right) \widetilde{%
\varepsilon }_{h}^{Q}+\eta _{0}x_{0}\right] }  \label{sig_AL^Q}
\end{equation}

Note that in deriving Eqs.(\ref{sig_AL^Q}), (\ref{SCFeq_Q}) and (\ref%
{sig_DOS^Q}), with the field independent parameter $\eta _{0}$, the small
diamagnetic energy term was neglected.

\section{Robustness of the fitting process}

In this appendix we present results (see Fig.3) of a fitting process similar
to that presented in Ref.\cite{MZPRB2021}, in which the prefactor of the AL
conductivity term is 1/2 of that used in Ref.\cite{MZPRB2021}. The level of
agreement between these calculations and the experimental data is preserved
if the values of the dimensionless spin-orbit scattering parameter used in
Ref.\cite{MZPRB2021}, i.e. $\beta _{0}=14$ ($R_{N}=7.5$ k$\Omega $) and $%
\beta _{0}=11$ ($R_{N}=20.5$ k$\Omega $), are changed in the new fitting to $%
\beta _{0}=16$ and $\beta _{0}=12$ respectively. The phenomenological
parameters determining the quantum tunneling attempt rates and the normal
state conductivity should also slightly modified. All the other parameters
of microscopic origins can remain fixed.


\begin{thebibliography}{99}
\bibitem{MZAnalCont} T. Maniv and V. Zhuravlev, Unpublished results. It is
shown that for frequencies in the region $\Omega <2\pi k_{B}T/\hbar $ the
analytic continuation $i\Omega _{\nu }\rightarrow \Omega \rightarrow 0$
directly from the discrete Matsubara sum Eq.(\ref{Q_AL}) is by a factor of 2
larger than that obtained by using the common contour-integration scheme
consisting of three sub-contours as done, e.g. in \cite{LV05}. Employing the
contour-integration scheme for $i\Omega _{\nu }\rightarrow \Omega
\rightarrow 0$ with only two sub-contours, i.e. by excluding the
intermediate sub-contour, one recovers the result obtained within the direct
analytic continuation.
\bibitem{footnote2} Estimation of the argument of the logarithmic factor in
Eq.(\ref{sig^fluctdt0}) just above the "nominal" critical field $%
H_{c\parallel 0}^{\ast }=4.5$T ($\varepsilon _{h\gtrsim 1}=0.05$) in the $%
t\rightarrow 0$ limit, based on typical values of our fitting parameters
yields $\left( \eta _{0}\left( h\right) x_{0}/h^{2}\varepsilon _{h}\right)
_{h\gtrsim 1}\approx 1.3$. \ Estimations of the field-dependent prefactors
of the AL and the DOS conductivities under the same conditions yield,
respectively: $\left( \eta _{0}\left( h\right) /4h^{2}\right) _{h=1}\simeq
\left( 3.5\zeta \left( 3\right) h^{2}/\eta _{0}\left( h\right) \right)
_{h=1}\simeq 1$.
\end{thebibliography}
\end{document}